\documentclass[10pt]{article}

\usepackage{graphicx}%
\usepackage{amsmath,amssymb,amsthm,upref,bm}%
\usepackage{labelfig}%
\usepackage{mathrsfs}

\DeclareMathAlphabet{\varmathbb}{U}{pxsyb}{m}{n}

\newcommand{\MF}[1]{\mathop{#1\vrule height0.45ex width0pt}\nolimits}
 
\newcommand{\D}{\mathrm{d}\kern0.2pt}%
\newcommand{\ii}{\kern0.05em\mathrm{i}\kern0.05em}%
\newcommand{\E}[1]{\textrm{e}^{#1}}%
\renewcommand{\vec}[1]{\bm{#1}}%
\newcommand{\RR}{\varmathbb{R}}%
\newcommand{\Cb}{\varmathbb{C}}%

\newcommand{\Yfunc}{\mathcal{Y}} 

\def\transp{\mathsf{T}}

\begin{document}

\baselineskip=4.4mm

\makeatletter

\title{\bf Freely floating structures trapping time-harmonic water waves
(revisited)}

\author{Nikolay Kuznetsov and Oleg Motygin}

\date{}

\maketitle

\vspace{-8mm}

\begin{center}
Laboratory for Mathematical Modelling of Wave Phenomena, \\ Institute for Problems
in Mechanical Engineering, Russian Academy of Sciences, \\ V.O., Bol'shoy pr. 61,
St. Petersburg 199178, Russian Federation \\ E-mail: nikolay.g.kuznetsov@gmail.com;
o.v.motygin@gmail.com
\end{center}

\begin{abstract}
We study the coupled small-amplitude motion of the mechanical system consisting of
infinitely deep water and a structure immersed in it. The former is bounded above by
a free surface, whereas the latter is formed by an arbitrary finite number of
surface-piercing bodies floating freely. The mathematical model of time-harmonic
motion is a spectral problem in which the frequency of oscillations serves as the
spectral parameter. It is proved that there exist axisymmetric structures consisting
of $N \geq 2$ bodies; every structure has the following properties: (i) a
time-harmonic wave mode is trapped by it; (ii) some of its bodies (may be none) are
motionless, whereas the rest of the bodies (may be none) are heaving at the same
frequency as water. The construction of these structures is based on a
generalization of the semi-inverse procedure applied earlier for obtaining trapping
bodies that are motionless although float freely.
\end{abstract}

\setcounter{equation}{0}

\footnotetext{The authors were supported by the Presidium of the Russian Academy of
Sciences (Programme $44 \Pi$: ``Fundamental Scientific Search Studies of
Perspectives for Development of the Arctic Region'').}

\section{Introduction}

This paper deals with the coupled problem describing the time-harmonic motion of the
mechanical system that consists of an inviscid, incompressible, heavy fluid (water)
in which a partially immersed structure floats freely. The latter means that there
are no external forces acting on it other than gravity (for example, due to
constraints on its motion). It is assumed that water occupies (together with the
immersed part of the structure) a half-space and the structure consists of a finite
number $N \geq 2$ of bounded surface-piercing bodies. The water motion is supposed
to be irrotational and the surface tension is neglected on the free surface of
water; moreover, the motion of the system is supposed to be of small amplitude near
equilibrium, which allows us to use a linear model. In our previous papers \cite{KM}
and \cite{KM1}, we considered the case of a single body which floats freely in water
of finite and infinite depth, respectively.

In his pioneering article \cite{John1}, F. John developed the time-dependent model
for a single freely floating body. The two-dimensional version of the coupling
conditions proposed by him (in particular, the equations of body's motion) were
presented in the convenient matrix form in \cite{NGK10}. In \cite{VidNaz}, a similar
form was developed for the three-dimensional problem and here we use its
generalisation to the case of multiple bodies. Assuming that the water motion is
simple harmonic in time, we reduce the time-dependent problem to a coupled spectral
problem with the frequency of oscillations playing the role of spectral parameter;
it appears in the boundary conditions, the radiation condition at infinity as well
as in the equations of motion of each body. Since the system's energy is proved to
be finite, it is possible to reformulate this spectral problem as an operator
equation in a Hilbert space (see \cite{VidNaz}, where this approach was developed
for a similar problem in a channel symmetric about its centre-plane). However, such
a formulation is superfluous for our purpose of constructing trapping structures and
the corresponding trapped modes.

In our papers \cite{KM} and \cite{KM1}, various uniqueness theorems were proved for
the time-harmonic problem for finite and infinite depth of water, respectively. In
the first of these papers, the original proof of John \cite{John2} was essentially
simplified. For proving the uniqueness theorem in \cite{KM1}, a new non-dimensional
form of the problem was proposed; this allowed us to evaluate the lower bound for
frequencies at which the uniqueness is guaranteed. The last paper also contains
various examples of a single body with the following properties. It is motionless,
floats freely and traps axisymmetric wave modes. Earlier, similar examples were
obtained for the two-dimensional problem in \cite{NGK10}.

One has to keep in mind that in the vast majority of published papers the questions
of uniqueness and existence of solutions and of trapped modes are studied for the
scattering and radiation problems under the assumption that an immersed body or
several such bodies are fixed (see \cite{LWW} and \cite{LM} for surveys, whereas the
most recent uniqueness theorems can be found in \cite{NGK08} and \cite{MoM}). Before
2010, the only rigorous result for the problem of a freely floating body was that of
John \cite{John2}, who proved a uniqueness theorem (unfortunately, without
formulating the problem explicitly). No other rigorous results about this problem
had been obtained until recently. However, after 2005 a number of authors considered
the question of trapped modes at the heuristic level and various two-dimensional and
axisymmetric trapping structures were proposed by virtue of numerical computations
(see \cite{MM,MM1,EP,N,PE1,PE2,FM}, which are listed in the chronological order). In
most of these papers, a simplified model is treated; it deals with a freely floating
body constrained to the heave motion only. The exception is the article \cite{PE2}
in which an example of trapping structure is considered whose motion is combined
(heave and sway).

In the present paper, our aim is to construct explicitly trapped modes, that is,
eigensolutions of the coupled time-harmonic problem involving freely floating
structures that consist of multiple bodies; some of these (may be none) are
motionless, whereas the rest part of bodies (may be none) are in the heave motion.
For this purpose we apply the so-called inverse method that replaces finding a
solution to a problem in a given domain by determining a physically acceptable water
region for a given solution. It is worth mentioning that this method widely used in
continuum mechanics prior to the advent of computers (see \cite{Nem} for a survey)
appears in two forms distinguished by the involvement of boundary conditions. If
some of these conditions, but not all, are prescribed at the outset, the method is
referred to as semi-inverse and this particular form of it is used here.

The paper's plan is as follows. We begin with formulating the time-dependent problem
in \S\,2.1. Then we apply an ansatz that introduces complex-valued unknowns
appropriate for considering time-harmonic oscillations and reduces the
time-dependent problem to a coupled spectral problem (\S\,2.2). The brief \S\,3
deals with the energy of the coupled time-harmonic motion; it also contains the
definition of a trapped mode. In the main \S\,4, various trapped modes and the
corresponding axisymmetric water domains are constructed.

\section{Statement of the problem}

\begin{figure}
\centering
 \SetLabels
 \L (0.05*0.91) {\small $g$}\\
 \L (0.96*0.62) {\small $x_2$}\\
 \L (0.517*0.95) {\small $y$}\\
 \L (0.487*0.515) {\small $x_1$}\\
 \L (0.3*0.42) {\small $B_k$}\\
 \L (0.7*0.51) {\small $B_k$}\\
 \L (0.367*0.545) {\small $\widehat{B}_k$}\\
 \L (0.54*0.28) {\small $W$}\\
 \L (0.08*0.605) {\small $F$}\\
 \L (0.81*0.76) {\small $F$}\\
 \L (0.85*0.605) {\small $F$}\\
 \endSetLabels
 \mbox{\kern-1mm}\AffixLabels{\includegraphics[width=64mm]{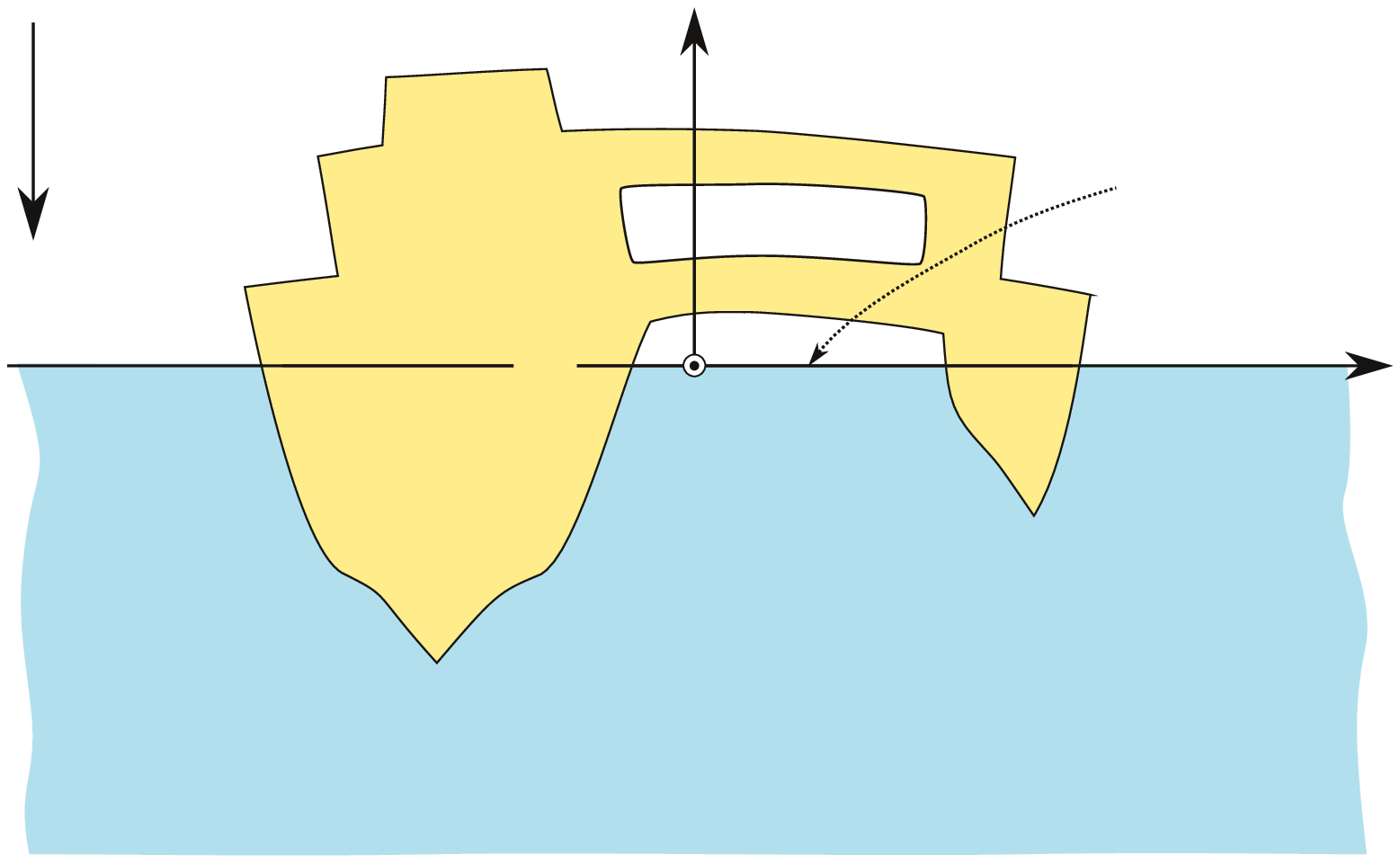}}
 \vspace{-4mm}
 \caption{A body with two immersed parts.}
 \vspace{1mm}
 \label{fig1}
\centering
 \SetLabels
 \L (0.96*0.62) {\small $x_2$}\\
 \L (0.517*0.95) {\small $y$}\\
 \L (0.487*0.515) {\small $x_1$}\\
 \L (0.26*0.42) {\small $B_1$}\\
 \L (0.69*0.46) {\small $B_2$}\\
 \L (0.31*0.545) {\small $\widehat{B}_1$}\\
 \L (0.645*0.545) {\small $\widehat{B}_2$}\\
 \L (0.02*0.72) {\small $D_1$}\\
 \L (0.84*0.72) {\small $D_2$}\\
 \L (0.54*0.28) {\small $W$}\\
 \L (0.18*0.28) {\small $S_1$}\\
 \L (0.75*0.38) {\small $S_2$}\\
 \L (0.06*0.6) {\small $F$}\\
 \L (0.54*0.6) {\small $F$}\\
 \L (0.88*0.6) {\small $F$}\\
 \endSetLabels
 \mbox{\kern-1mm}\AffixLabels{\includegraphics[width=64mm]{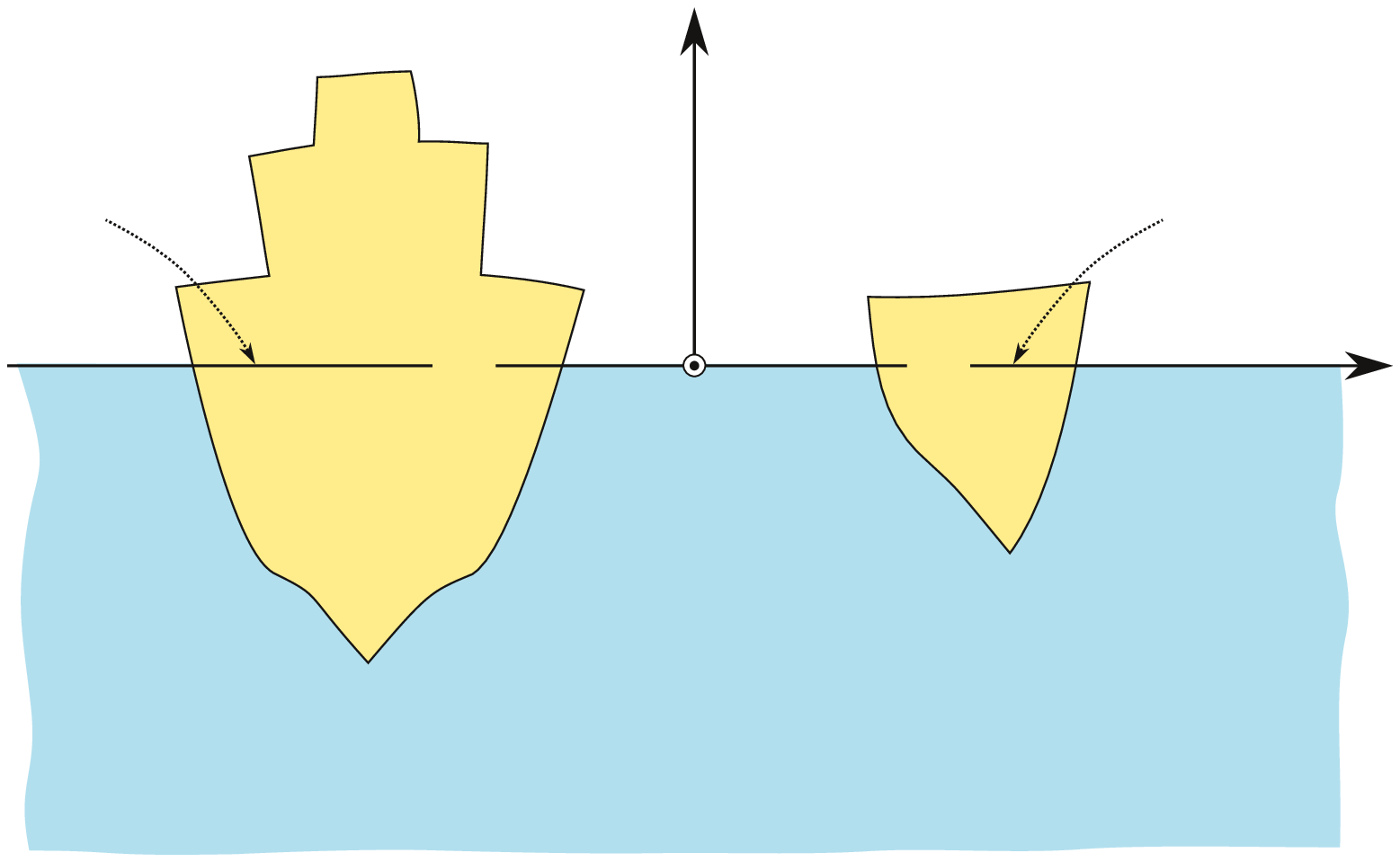}}
 \vspace{-4mm}
 \caption{Definition sketch for two bodies.}
 \vspace{-2mm}
 \label{fig2}
\end{figure}

Let the Cartesian coordinates $(\vec{x},y)$, $\vec{x}=(x_1,x_2)$, be such that the
$y$-axis is directed upwards, whereas the mean free surface of water lies in the
$\vec{x}$-plane, and so the water domain is a subset of $\RR^3_- = \{ \vec{x} \in
\RR^2, \, y<0 \}$. The domain occupied by the $k$th body in its equilibrium position
we denote by $\widehat{B}_k$, $k=1,\dots,N$; its immersed part $B_k = \widehat{B}_k
\cap \RR^3_- \neq \emptyset$ can consist of several connected components (see
fig.~\ref{fig1}). Let $B = \cup_{k=1}^N B_k$ and $W = \RR^3_- \setminus
\skew2\overline{B}$ denote the structure's submerged part and the water domain,
respectively. It is supposed that $W$ is simply connected, whereas $B$ has at least
$N \geq 2$ connected components (see fig.~\ref{fig2}); their number is greater than
$N$ if bodies like that shown in fig.~\ref{fig1} are present. Furthermore, $S_k =
\partial B_k \cap \RR^3_-$ and $F = \{ y \nobreak = \nobreak 0 \} \setminus
\skew2(\cup_{k=1}^N \overline{D}_k)$ stand for the wetted surface of the $k$th body
and the free surface of water in its mean position, respectively; here $D_k =
\widehat{B}_k \cap \{ y \nobreak = \nobreak 0 \}$ (see fig.~2); for the sake of
brevity we put $S = \cup_{k=1}^N S_k$.

\vspace{-2mm}

\subsection{Time-dependent problem}

\looseness=-1 In the linearised time-dependent setting for $N=1$ obtained in
\cite{John1}, the coupled motion is described in terms of the so-called first-order
unknowns: the velocity potential for the water motion and a vector characterising
the motion of the body's centre of mass. Here we give the problem's statement for $N
\geq 2$ on the basis of the assumptions listed above which are the same as in
\cite{John1}.

The first-order velocity potential $\Phi (\vec{x},y;t)$ exists because the water
motion is supposed to be irrotational and $W$ is simply connected. Hence the
velocity field is equal to $\nabla \Phi (\vec{x},y;t)$ ($\nabla =
(\partial_{x_1}, \partial_{x_2}, \partial_{y})$ is the spacial gradient), and
the continuity equation takes the form: 
\begin{equation}
\nabla^2 \Phi = 0\quad \mbox{in}\ W \quad \mbox{for\ all}\ t. \label{lap}
\end{equation}
In order to specify the behaviour of $\Phi$ near $\partial W$, we require that this
function belongs to the Sobolev class $H^1_{\mathrm{loc}} (W)$. The standard linear
boundary condition on the free surface has the following form:
\begin{equation}
\partial_{tt} \Phi + g \partial_{y} \Phi = 0 \quad \mbox{on}\ F \quad \mbox{for all}
\ t . \label{tt}
\end{equation}
Here $g > 0$ is the acceleration due to gravity that acts in the direction opposite
to the $y$-axis (see fig.~\ref{fig1}). Equality \eqref{tt} is a consequence of
Bernoulli's equation and the kinematic condition both taken linearised on the mean
free surface; the first of these conditions expresses the fact that the pressure is
constant on $F$, whilst the second one means that there is no transfer of matter
across $F$.

Furthermore, the following set of vectors $\vec{q}^{(k)} (t) \in \RR^6$
characterises the motion of the centre of mass of the $k$th body about its given
equilibrium position $\bigl( \vec{x}^{(k)}_0, y^{(k)}_0 \bigr)$; namely, for every
$k=1,\dots,N$

\vspace{1mm}

\noindent $\bullet$ the horizontal and vertical displacements are $q_1^{(k)}$,
$q_2^{(k)}$ and $q_4^{(k)}$, respectively;

\noindent $\bullet$ $q_3^{(k)}$ and $q_5^{(k)}$, $q_6^{(k)}$ are the angles of
rotation about the axes that go through $\bigl( \vec{x}^{(k)}_0, y^{(k)}_0 \bigr)$
parallel to the $y$- and $x_1$-, $x_2$-axes, respectively. 

\vspace{1mm}

\noindent The following kinematic condition couples $\Phi$ and $\vec{q}^{(k)}$
for each body: 
\begin{equation}
 \partial_{\vec{n}} \Phi (\vec{x},y;t) = [\vec{n} (\vec{x},y)]^\transp
 \bm{D}^{(k)}_0 (\vec{x},y) \, \Dot{\vec{q}}^{(k)} (t) \quad \mbox{on}\ S_k \quad
 \mbox{for all} \ t .
 \label{eq:onB}
\end{equation}
By $^\transp$ the operation of matrix transposition is denoted (a vector is
considered as a one-column matrix) and $\vec{n}$ is the unit normal to $\partial W$
(in particular, to $S_k$) directed to the exterior of $W$. The vector
$\Dot{\vec{q}}$ (the dot stands for the time derivative) characterises the body
motion in the following manner: $(\Dot q_1, \Dot q_2, \Dot q_4)^\transp$ is the
velocity vector of the translational motion and $(\Dot q_3, \Dot q_5, \Dot
q_6)^\transp$ is the vector of angular velocities. The $3\!\times\!6$ matrix
$\bm{D}^{(k)}_0 (\vec{x},y)$ is defined as follows: \\[-3mm]
\[ \bm{D}^{(k)}_0 (\vec{x},y) = \bm{D} \bigl( \vec{x} - \vec{x}^{(k)}_0, y -
 y^{(k)}_0 \bigr) , \quad \mbox{where} \ \ \bm{D} (\vec{x},y) = \left[
\begin{matrix}
 1 & 0 & x_2 & 0 & 0 & -y
\\
 0 & 1 & -x_1 & 0 & y & 0
\\
 0 & 0 & 0 & 1 & -x_2 & x_1 
\end{matrix}
 \right] .
\]
The latter describes the motion of a rigid body so that its elements are in
conformity with the order of components of the corresponding vector~$\vec{q}$.

The linearised system of equations describing the motion of the $k$th body
expresses the conservation of its linear and angular momentum. In the absence of
external forces other than gravity this system is as follows: 
\begin{equation}
 \bm{E}^{(k)}_0 \Ddot{\vec{q}}^{(k)} (t) = - \int_{S_k} \partial_{t} \Phi (\vec{x},y;t)
 [\bm{D}^{(k)}_0 (\vec{x},y)]^\transp \vec{n} (\vec{x},y) \, \D{}s
 - g \bm{K}^{(k)}_0 \vec{q}^{(k)} (t) \ \ \mbox{for all} \ t . 
 \label{eq:moveq}
\end{equation}
Here $\Ddot{\vec{q}}^{(k)} (t)$ is the acceleration vector of the $k$th body and
$\bm{E}^{(k)}_0$ is its mass/inertia matrix defined as follows:
\begin{equation*}
 \bm{E}^{(k)}_0 = \rho_0^{-1} 
 \int_{\widehat{B}_k} \rho_k (\vec{x},y) [\bm{D}^{(k)}_0 (\vec{x},y)]^\transp
 \bm{D}^{(k)}_0 (\vec{x},y) \, \D \vec{x} \, \D{}y ,
\end{equation*}
where $\rho_k (\vec{x},y) \geq 0$ is the density distribution within the $k$th body
and $\rho_0 > 0$ is the constant density of water. A direct calculation gives the
following explicit form of this matrix:
\begin{equation}
\bm{E}^{(k)}_0 =
\begin{pmatrix}
 I^{\widehat{B}_k} & 0 & 0 & 0 & 0 & 0 \\
 0 & I^{\widehat{B}_k} & 0 & 0 & 0 & 0 \\
 0 & 0 & I^{\widehat{B}_k}_{x_1x_1} + I^{\widehat{B}_k}_{x_2x_2} & 0 
 & -I^{\widehat{B}_k}_{x_1y} & - I^{\widehat{B}_k}_{x_2y} \\
 0 & 0 & 0 & I^{\widehat{B}_k} & 0 & 0 \\
 0 & 0 & - I^{\widehat{B}_k}_{x_1y} & 0 & I^{\widehat{B}_k}_{x_2x_2} +
 I^{\widehat{B}_k}_{yy} & - I^{\widehat{B}_k}_{x_1x_2} \\
 0 & 0 & - I^{\widehat{B}_k}_{x_2y} & 0 & - I^{\widehat{B}_k}_{x_1x_2} &
  I^{\widehat{B}_k}_{x_1x_1} + I^{\widehat{B}_k}_{yy}
\end{pmatrix} .
\label{eq:Efull}
\end{equation}
The matrix elements are composed of various moments of the whole body
$\widehat{B}_k$; namely, 
\begin{equation*}
 I^{\widehat{B}_k} = \rho_0^{-1} \int_{\widehat{B}_k} \rho_k (\vec{x},y) \, 
 \D \vec{x} \, \D{}y , \quad I^{\widehat{B}_k}_{\sigma \tau} = \rho_0^{-1}
 \int_{\widehat{B}_k} \rho_k (\vec{x},y) \, \bigl( \sigma - \sigma^{(k)}_0 \bigr) \,
 \bigl( \tau - \tau^{(k)}_0 \bigr) \, \D \vec{x} \, \D{}y . 
\end{equation*}
Here $\sigma$ and $\tau$ stand for the corresponding coordinates, whereas
$\sigma^{(k)}_0$ and $\tau^{(k)}_0$ are taken equal to $x_{1 0}^{(k)}, x_{2
0}^{(k)}$ and $y^{(k)}_0$ so that these coordinates are the same as $\sigma$ and
$\tau$, respectively. In formula \eqref{eq:Efull}, it is taken into account that
for any coordinate $\sigma$ we have 
\[ I^{\widehat{B}_k}_\sigma = \rho_0^{-1} \int_{\widehat{B}_k} \rho_k (\vec{x},y) \, 
\bigl( \sigma - \sigma^{(k)}_0 \bigr) \, \D \vec{x} \, \D{}y = 0 ,
\]
which is a consequence of the definition of $\bigl( \vec{x}^{(k)}_0, y^{(k)}_0
\bigr)$. It is obvious that every matrix $\bm{E}^{(k)}_0$ is symmetric; moreover, it
is straightforward to verify that all these matrices are positive definite.

In the right-hand side of \eqref{eq:moveq}, we have forces and their moments;
namely, the first term is of the hydrodynamic origin, whereas the second one is
related to the buoyancy (see, for example, \cite{John1} and \cite{Mei}). The
matrix in the second term has the following blockwise form: 
\begin{equation}
 \bm{K}^{(k)}_0 =
\begin{pmatrix}
 \mathbb{O}_3 & \mathbb{O}_3 \\
 \mathbb{O}_3 & \bm{\widehat K}^{(k)}_0
\end{pmatrix} ,
 \quad \mbox{where} \ \ \bm{\widehat K}^{(k)}_0 =
\begin{pmatrix}
 I^{D_k} & I^{D_k}_{2} & - I^{D_k}_{1} \\
 I^{D_k}_{2} & I^{D_k}_{22} + I^{B_k}_y & - I^{D_k}_{12} \\
 - I^{D_k}_{1} & - I^{D_k}_{12} & I^{D_k}_{11} + I^{B_k}_y 
\end{pmatrix}
\label{eq:KK'}
\end{equation}
and $\mathbb{O}_3$ is the null $3\!\times\!3$ matrix. The elements of
$\bm{\widehat K}^{(k)}_0$ involve the following moments: 
\begin{equation*}
\begin{gathered}
  I^{D_k} = \int_{D_k} \D \vec{x} , \quad
  I^{B_k}_y = \int_{B_k} \bigl( y - y^{(k)}_0 \bigr) \, \D \vec{x} \, \D{}y , \\
  I^{D_k}_i = \int_{D_k} \bigl( x_i - x_{i 0}^{(k)} \bigr) \, \D \vec{x} , \quad
  I^{D_k}_{ij} = \int_{D_k} \bigl( x_i - x_{i 0}^{(k)} \bigr) \, \bigl( x_j 
  - x_{j 0}^{(k)} \bigr) \, \D \vec{x} , \quad i,j=1,2 .
\end{gathered}
\end{equation*}
It is clear that the matrix $\bm{K}^{(k)}_0$ is symmetric.

\vspace{1mm}

Relations (\ref{lap})--(\ref{eq:moveq}) must be augmented by the following
subsidiary conditions gua\-ranteeing equilibrium for each of $N$ floating bodies
and its stability:

\vspace{1mm}

\noindent $\bullet \ I^{\widehat{B}_k} = \int_{B_k} \D \vec{x} \, \D{}y$; that is,
the mass of water displaced by the $k$th body is equal to its own mass (Archimedes'
law).

\noindent $\bullet \ \int_{B_k} \bigl( x_i - x_{i 0}^{(k)} \bigr) \, \D \vec{x} \,
\D{}y = 0 ,\ i=1,2$; that is, the center of buoyancy of the $k$th body lies on the
same vertical line as its centre of mass (see, for example, \cite[\S\,8.2.3]{Mei}).

\noindent $\bullet$ Every matrix $\bm{K}^{(k)}_0$ is positive semi-definite, whereas
$\bm{\widehat K}^{(k)}_0$ is positive definite. This is the classical condition
yielding the stability of the equilibrium position of the $k$th body (see, for
example, \cite[\S\,2.4]{John1}). As usual, the stability means that an
instantaneous, infinitesimal disturbance causes the position changes remaining
infinitesimal for all subsequent times with the exception of purely horizontal
motion.

\vspace{1mm}

Of course, relations \eqref{lap}--\eqref{eq:moveq} must be complemented by proper
initial conditions in order to obtain a well-posed initial-value problem (see
\cite{Beale}, \S\,3, where this question is considered). However, our aim is to
study free time-harmonic oscillations of the system not depending on the initial
conditions. A heuristic explanation how this phenomenon is to be conceived is
discussed in detail in \cite{John2}, p.~46.

\subsection{Time-harmonic problem}

In order to formulate the problem of coupled time-harmonic motion for the
mechanical system described in \S\,2.1, we assume $\omega > 0$ to be the radian
frequency of oscillations and represent the velocity potential and the
displacement vectors in the following form: 
\begin{equation*}
 \bigl( \Phi (\vec{x},y;t), \vec{q}^{(1)} (t), \dots, \vec{q}^{(N)} (t) \bigr) =
 \Re \bigl\{ \E{-\ii \omega t} \bigl( \varphi (\vec{x},y), \ii \vec{\chi}^{(1)}, \dots,
 \ii \vec{\chi}^{(N)} \bigr) \bigr\} .
\end{equation*}
Here $\varphi$ is a complex-valued function and $\vec{\chi}^{(k)} \in \Cb^6$.
Substituting the latter expression into relations \eqref{lap}--\eqref{eq:moveq},
we immediately get 
\begin{gather}
 \nabla^2 \varphi = 0 \quad \mbox{in} \ W ,
 \label{eq:1} \\
 \partial_{y} \varphi - \nu \varphi = 0 \ \ \mbox{on} \ F , \quad \nu = 
 \frac{\omega^2}{g} \, ,
 \label{eq:2} \\
 \partial_{\vec{n}} \varphi = \omega \, \vec{n}^\transp\! \bm{D}^{(k)}_0
 \vec{\chi}^{(k)} \quad \mbox{on} \ S_k \, , \label{eq:4} \\
 \omega^2 \bm{E}^{(k)}_0 \vec{\chi}^{(k)} = - \omega \int_{S_k} \varphi 
 [\bm{D}^{(k)}_0 (\vec{x},y)]^\transp \vec{n} \, \D{}s
 + g \, \bm{K}^{(k)}_0 \vec{\chi}^{(k)} ,
 \label{eq:5}
\end{gather}
where the last two conditions must hold for every $k=1,\dots,N$. We specify the
behaviour of $\varphi$ at infinity so that the velocity potential $\Phi$ describes
outgoing waves by imposing the radiation condition
\begin{equation}
 \int_{W\cap\{|\vec{x}|=a\}} \bigl| \partial_{|\vec{x}|} \varphi - \ii \nu
 \varphi \bigr|^2 \, \D{}s = o (1) \quad \mbox{as} \ a \to \infty .
 \label{eq:6}
\end{equation}
It is natural that this condition is the same as in the water-wave problem for
fixed obstacles (see, for example, \cite{John2}). In the boundary value problem
\eqref{eq:1}--\eqref{eq:6}, $\omega$ is a spectral parameter sought together with
the eigenvector $\big( \varphi, \vec{\chi}^{(1)}, \dots, \vec{\chi}^{(N)} \big)$.

Since, generally speaking, $\partial W$ is not smooth and we assumed that $\varphi
\in H^1_{\mathrm{loc}} (W)$, it is sufficient to understand the whole set of
relations \eqref{eq:1}--\eqref{eq:4} in the sense of the integral identity
\begin{equation}
 \int_{W} \nabla \varphi \nabla \psi \,\D \vec{x}\, \D{}y = \nu \int_{F} 
 \varphi \, \psi \, \D \vec{x} + \omega \sum_{k=1}^N \int_{S_k} \psi \,
 \vec{n}^\transp\! \bm{D}^{(k)}_0 \vec{\chi}^{(k)} \, \D{}s ,
\label{eq:intid}
\end{equation}
which must hold for all smooth functions $\psi$ having a compact support in
$\overline W$.

\section{On the energy of the coupled \\ time-harmonic motion}

In \cite{KM} and \cite{KM1}, the following intuitively `obvious' assertion was
proved for a single body immersed in water of finite and infinite depth,
respectively. If there is no net input of energy into the motion of water or bodies
forming the structure, then there is no wave radiation to infinity, and so the total
energy of the coupled motion is finite. In this section, we generalise this fact to
the case of a structure consisting of multiple bodies, thus showing that a
non-trivial solution of problem (\ref{eq:1})--(\ref{eq:5}) and (\ref{eq:6})
describes waves trapped by the structure.

It is known (see, for example, \cite[\S\,2.2.1]{LWW}) that if $\varphi$ satisfies
relations (\ref{eq:1}), (\ref{eq:2}) and (\ref{eq:6}), then it has the following
asymptotic representation at infinity:
\begin{equation}
 \varphi (\vec{x},y) = A (\theta) |\vec{x}|^{-1/2} \E{\nu (y + \ii |\vec{x}|)} + 
 R (\vec{x},y) .
\label{eq:uas}
\end{equation}
Moreover, the remainder behaves as follows:
\begin{equation}
 |R| , \, |\nabla R| = O \Bigl( \E{\nu y} (1 + \nu |\vec{x}|)^{-3/2} +
 \left[ \nu^2 (|\vec{x}|^{2} + y^2) \right]^{-1} \Bigr) \ \ \mbox{as} \ \ 
 \nu^2 (|\vec{x}|^{2} + y^2) \to \infty ,
\label{eq:rem}
\end{equation}
and the equality
\begin{equation}
 \frac12 \int_0^{2\pi} \bigl|A (\theta)\bigr|^2 \, \D{}\theta =
 -\Im \int_S \overline{\varphi}\,\partial_{\vec{n}} \varphi \,\D{}s 
\label{eq:ener}
\end{equation}
holds for the coefficient in the leading term. Here $\theta$ is the polar angle in
the $(x_1, x_2)$-plane measured anti-clockwise and we recall that $S$ denotes the
union of all $S_k$.

In the same way as in \cite{KM,KM1} it follows from (\ref{eq:4}), (\ref{eq:5}) and
(\ref{eq:ener}) that
\begin{equation*}
 \frac12 \int_0^{2\pi} \bigl|A (\theta)\bigr|^2 \, \D{}\theta = \Im \sum_{k=1}^N
 \Bigl\{ \omega^2 \, \overline{\vec{\chi}^{(k)}}^\transp \bm{E}^{(k)}_0 \vec{\chi}^{(k)}
 - g \overline{\vec{\chi}^{(k)}}^\transp \bm{K}^{(k)}_0 \vec{\chi}^{(k)} \Bigr\} .
\end{equation*}
Since the expression in braces is real, we get that $A(\theta) = 0$, and so the
behaviour of $\varphi$ at infinity is given by formula (\ref{eq:rem}). Combining
this fact and the assumption that $\varphi \in H^1_{\mathrm{loc}} (W)$, we arrive at
the following assertion.

\vspace{1mm}

\noindent {\sc Lemma 3.1.} {\it Let\/ $\big( \varphi, \vec{\chi}^{(1)}, \dots,
\vec{\chi}^{(N)} \big)$ be a solution of problem\/ $(\ref{eq:1})$--$(\eqref{eq:5})$
and\/ $(\ref{eq:6})$, then the first component\/ $\varphi$ belongs to the Sobolev
space\/ $H^1 (W)$.}

\vspace{1mm}

Now we are in a position to extend Proposition~1 in \cite{KM1} to the case of
structures consisting of multiple bodies.

\vspace{1mm}

\noindent {\sc Proposition 3.2.} {\it Let\/ $\big( \varphi, \vec{\chi}^{(1)}, \dots,
\vec{\chi}^{(N)} \big)$ be a solution of problem\/ $(\ref{eq:1})$--$(\ref{eq:6})$, then
\begin{equation}
 \int_W |\nabla \varphi|^2 \, \D{} \vec{x} \, \D{}y < \infty \quad \mbox{and} \quad 
 \int_F |\varphi|^2 \, \D{} \vec{x} < \infty ,
\label{eq:finenerg}
\end{equation}
that is, the kinetic and potential energy of the water motion are finite.
Moreover, the following equality holds
\begin{equation}
 \int_W |\nabla\varphi|^2 \, \D{} \vec{x} \, \D{}y + \omega^2 \sum_{k=1}^N
 \overline{\vec{\chi}^{(k)}}^\transp \bm{E}^{(k)}_0 \vec{\chi}^{(k)} = \nu \int_F 
 |\varphi|^2 \,\D{} \vec{x} + g \, \overline{\vec{\chi}^{(k)}}^\transp 
 \bm{K}^{(k)}_0 \vec{\chi}^{(k)} ,
\label{eq:equipart}
\end{equation}
thus expressing the equipartition of energy of the coupled motion.}

\vspace{1mm}

\noindent {\it Proof.} Relations (\ref{eq:finenerg}) are an immediate consequence of
Lemma~3.1. For proving equality (\ref{eq:equipart}) we introduce an infinitely
differentiable cut-off function $\zeta_a (|\vec{x}|, y)$ equal to one on $\{
|\vec{x}| \leq a , -a \leq y \leq 0 \}$ and to zero when either $|\vec{x}| \geq a+1$
or $y \leq -a-1$. Let $a$ be so large that the wetted surface $S$ of the whole
structure lies within the truncated cylinder $\{ |\vec{x}| < a, y > -a \}$, then
substituting $\psi = \overline \varphi \, \zeta_a (|\vec{x}|)$ into
\eqref{eq:intid}, we see that relations (\ref{eq:uas}) allow us to let $a \to
\infty$, which combined with (\ref{eq:5}) gives the required equality.

\vspace{1mm}

These assertions show that if $\big( \varphi, \vec{\chi}^{(1)}, \dots,
\vec{\chi}^{(N)} \big)$ is a solution of problem (\ref{eq:1})--(\ref{eq:6}) with
complex-valued components, then its real and imaginary parts separately satisfy this
problem. This allows us to consider $\big( \varphi, \vec{\chi}^{(1)}, \dots,
\vec{\chi}^{(N)} \big)$ as an element of the real product space $H^1 (W) \times
\RR^{6 N}$ in what follows.

\vspace{1mm}

\noindent {\sc Definition 3.3.} Let the subsidiary conditions concerning the
equilibrium position (see \S~2) hold for a freely floating structure. A
non-trivial real solution 
\[ \big( \varphi, \vec{\chi}^{(1)}, \dots, \vec{\chi}^{(N)} \big) \in H^1 (W) 
\times \RR^{6 N}
\]
of problem (\ref{eq:1})--(\ref{eq:5}) is called a {\it mode trapped}\/ by the
structure, whereas the corresponding value of $\omega$ is referred to as a {\it
trapping frequency}.

\section{Trapped modes and the corresponding \\ axisymmetric trapping structures}

In this section, we construct trapped modes with axisymmetric velocity fields and
the corresponding structures consisting of arbitrarily large but fixed number $N$ of
axisymmetric bodies; a part of these (may be all or none) are motionless, whereas
the rest (may be none or all) are heaving. In order to find such bodies we modify
the semi-inverse procedure applied in \cite{KM} using not only a special choice of
the velocity potential, but also a particular form of the vectors
$\vec{\chi}^{(k)}$, $k = 1,\dots,N$. The potential is defined so that it satisfies
the Laplace equation and the free-surface boundary condition; moreover, it does not
radiate waves to infinity. Since a set of bodies with axisymmetric immersed parts is
sought, level surfaces of a Stokes stream function are used for finding admissible
wetted surfaces of motionless bodies. Besides, a special term is added to this
stream function in order to find admissible wetted surfaces of heaving bodies as
level surfaces of the modified stream function (another its modification was
proposed in \cite{MM1}).

\subsection{Velocity potentials of trapped modes}

Let us fix $\omega > 0$ arbitrarily, and this value will serve as the trapping
frequency. A trap\-ped mode is sought in the form $\bigl( \omega \nu^{-2} \varphi_*,
\bm{d} \, \vec{\chi}^{(1)}_*, \dots, \bm{d} \, \vec{\chi}^{(N)}_* \bigr)$ with
dimensionless $\varphi_*$ and $\vec{\chi}^{(k)}_*$ and the diagonal matrix
$\mathrm{diag} \{\nu^{-1},\nu^{-1},1,\nu^{-1},1,1\}$ as $\bm{d}$. For the sake of
brevity, $\bigl( \varphi_*, \vec{\chi}^{(1)}_*, \dots, \vec{\chi}^{(N)}_* \bigr)$
will also be referred to as trapped mode.

Following the semi-inverse method, we define $\varphi_*$ explicitly for $y \leq 0$
(cf.~\cite{KM1},~\S\,3.1):
\begin{align}
\varphi_* (\nu |\vec{x}|, \nu y) ={}& 2 \int_0^\infty (k \cos k \nu y + \sin k
\nu y) I_0 (k \nu |\vec{x}|) K_1 (k \nu r) \frac{k^2\, \D k}{k^2 + 1}\notag \\
&{}- \pi^2 \E{\nu y} J_0 (\nu |\vec{x}|) Y_1 (\nu r) \quad  \mbox{when} \
|\vec{x}| < r ,
\label{eq:phi0<}
\\
\varphi_* (\nu |\vec{x}|, \nu y) ={}& 2 \int_0^\infty (k \cos k \nu y + \sin k \nu y)
K_0 (k \nu |\vec{x}|) I_1 (k \nu r) \frac{k^2\, \D k}{k^2 + 1} \notag \\ &{}- \pi^2
\E{\nu y} Y_0 (\nu |\vec{x}|) J_1 (\nu r) \quad \mbox{when} \ |\vec{x}| > r .
\label{eq:phi0>}
\end{align}
Here $r$ is a length to be specified below, $Y_0$ and $Y_1$ are the Neumann
functions of order zero and one, respectively, whereas $J_0$, $J_1$ and $I_0$,
$I_1$, $K_0$, $K_1$ denote the standard and modified Bessel functions of the
corresponding orders.

Using the asymptotic behaviour of $I_1$ and $K_1$ at infinity (see formulae
9.7.1 and 9.7.2 in \cite{AS} or \cite{W}, \S\S\,7.23,\,7.3), one immediately
obtains that both integrals diverge when $|\vec{x}| = r$ and $y=0$, and so
$\varphi_*$ has a singularity there. Furthermore, it is straightforward to
verify that $\varphi_*$ can be extended to the whole $\RR^3_-$ so that the
extension is harmonic. Moreover, the following boundary condition
\begin{equation}
 \partial_{y} \varphi_* - \nu \varphi_* = 0 \quad \mbox{holds on} \ \ \partial \RR^3_-
 \setminus \{ |\vec{x}| = r , \, y = 0 \} .
\label{eq:sing}
\end{equation}

Let us choose $r$ so that $\nu r = j_{1,m}$, where $j_{1,m}$ is one of the positive
zeros of $J_1$. (These zeros form an infinite sequence arranged in ascending order
as $m$ increases.) According to this choice of $r$, the second term in the
right-hand side of (\ref{eq:phi0>}) (it describes an outgoing wave) vanishes. In
what follows, we write $r_m$ for $\nu^{-1} j_{1,m}$ and denote by $\varphi_m$ the
function $\varphi_*$ with $r = r_m$. Therefore, $\varphi_m \in H^1 (W)$ for any
domain $W$ obtained by removing some neighbourhood of the circumference $\{
|\vec{x}| = r_m , \, y = 0 \}$ from $\RR^3_-$. Indeed, $\varphi_m$ satisfies the
same estimates at infinity as the remainder $R$ in (\ref{eq:uas}). Thus, any
function $\varphi_m$ defined by formulae (\ref{eq:phi0<}) and (\ref{eq:phi0>}) with
$r = r_m$ can serve as the first component of an eigensolution provided the vectors
$\vec{\chi}^{(k)}_*$ and the water domain $W$ are chosen properly.

\subsection{Stokes stream functions and streamlines}

In order to construct water domains we begin with introducing Stokes stream
functions that corresponds to the sequence $\varphi_m$. Namely, $\psi_m$ is
defined by virtue of the following relations:
\begin{equation}
\partial_{|\vec{x}|} \varphi_m = - (\nu |\vec{x}|)^{-1} \partial_{y} \psi_m , \quad
\partial_{y} \varphi_m = (\nu |\vec{x}|)^{-1} \partial_{|\vec{x}|} \psi_m .
\label{eq:ps}
\end{equation}
These equations give for $m=1,2,\dots$:
\begin{gather}
 \kern-27mm\psi_m (\nu |\vec{x}|, \nu y) = - \pi^2 \nu |\vec{x}| \, \E{\nu y} J_1
 (\nu |\vec{x}|) Y_1 (j_{1,m}) \nonumber \\[1mm] \ \ \ \ \ \ \ \ \ \ \ \ \ \ \ \ \ \
 \ \ \ - 2 \, \nu |\vec{x}| \Psi (\nu|\vec{x}|, \nu r_m, \nu y) \ \, \mbox{for}
 \ |\vec{x}| < r_m ,\; y \leq 0 , 
 \label{eq:psi0<} \! \\[1mm]
 \psi_m (\nu |\vec{x}|, \nu y) = -2 \, \nu  |\vec{x}| \Psi (\nu r_m, 
 \nu |\vec{x}|, \nu y) \ \ \, \mbox{for} \ |\vec{x}| > r_m , \ y \leq 0 ,
 \notag 
 \\[1mm]
 \mbox{where} \ \ \Psi (\sigma, \tau, \eta) = \int_0^\infty
 (k \sin k \eta - \cos k \eta) I_1 (k \sigma) K_1 (k \tau) \frac{k^2\, \D k}{k^2 + 1} \, .
 \nonumber
\end{gather}
The last function is defined for $(\sigma, \tau, \eta)$ such that $\eta \leq 0$,
$0 \leq \sigma \leq \tau$ and $\eta \neq 0$ when $\sigma = \tau$. The constant
of integration in this definition of $\psi_m$ is chosen so that $\psi_m (\nu
|\vec{x}|, \nu y) \to 0$ as $\nu^2 \bigl[ (|\vec{x}| -
 r_m)^2 + y^2 \bigr] \to \infty$.

Since the velocity field is axisymmetric, by a streamline we mean the curve in the
$(\nu |\vec{x}|, \nu y)$-plane given by the equation $\psi_m (\nu |\vec{x}|, \nu y)
= v$ with a constant $v$. (In fact, this equation defines axisymmetric surfaces and
streamlines are their vertical cross-sections.) First, we formulate some properties
of streamlines that will be used below.

\vspace{1mm}

\noindent {\sc Proposition 4.1.} (i) {\it In\/ $Q = \{ \nu |\vec{x}| > 0 , \, \nu y
< 0\}$, streamlines are smooth curves; their end-points belong to\/ $\partial Q$
for\/ $v \neq 0$ and to\/ $\partial Q \cup \{\infty\}$ for\/ $v=0$.} \\[1mm]
 (ii) {\it A streamline emanates from every point on the half-axis\/ $\{ \nu |\vec{x}| 
> 0 , \, \nu y = 0 \}$, except for the points, where\/ $\psi_m (\nu
|\vec{x}|,\nu y)$ attains its local extrema, and the point\/ $(\nu r_m, 0)$.}
\\[1mm]
 (iii) {\it For every $m \geq 1$ and all positive\/ $\nu$ and $v$ there exists 
a streamline such that $y=0$ at both its ends and the point\/ $(\nu r_m , 0)$
belongs to the segment connecting these end-points.}

\vspace{1mm}

This proposition is proved in \cite{KM1}, pp.~150--152.

\subsection{Motionless trapping structures}

\begin{figure}[t]
\begin{center}
 \SetLabels
 \L (-0.12*0.97) (a)\\
 \L (-0.12*0.61) (b)\\
 \L (0.9*0.725) {\small $\nu|\vec{x}|$}\\
 \L (0.9*-0.02) {\small $\nu|\vec{x}|$}\\
 \L (-0.04*0.49) {\small $\nu y$}\\
 \L (0.4*0.918) {\small $\psi_2 (\nu |\vec{x}|,0)$}\\
 \L (0.05*0.678) {\small Rigid shell}\\
 \L (0.44*0.678) {\small Air}\\
 \L (0.81*0.59) {\small Ballast}\\
 \endSetLabels
\leavevmode\kern2mm\AffixLabels{\includegraphics[width=64mm]{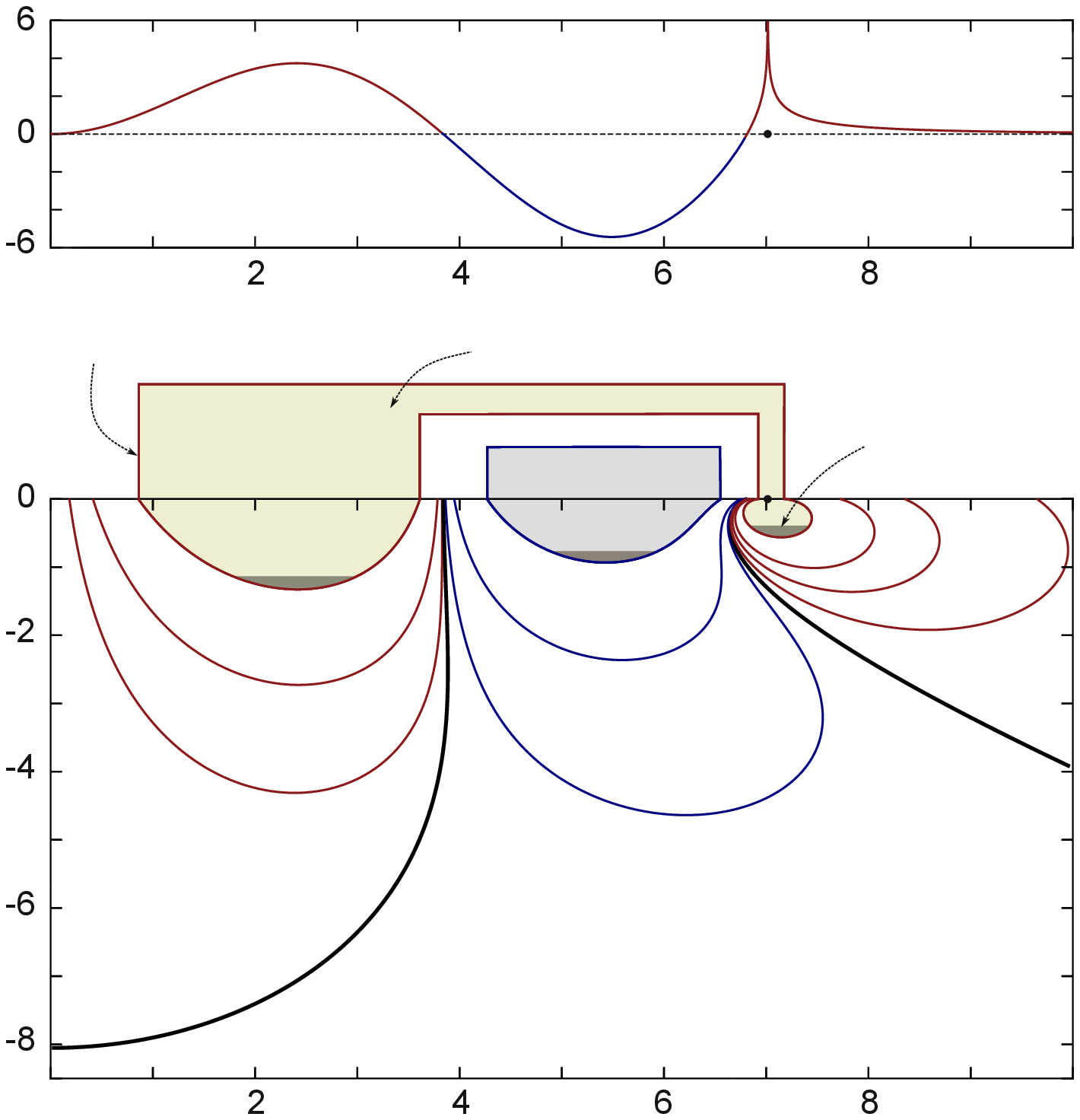}}
 \end{center}
 \vspace{-1mm}
 \caption{(a) The trace $\psi_2 (\nu |\vec{x}|,0)$. (b) Streamlines
 $\psi_2 (\nu |\vec{x}|, \nu y) = v$ for various values of $v$; nodal
 lines ($v=0$) are bold. Straight segments show how wetted surfaces are connected
 above the free surface to form 2 bodies. Darkly shaded layers show the ballast
 guaranteeing that the bodies are in equilibrium.}
 \label{fig:cont2}
\vspace{-2mm}
\end{figure}

It follows from (\ref{eq:ps}) that $\partial_{\vec{n}} \varphi_m$ vanishes on every
streamline of $\psi_m$, and so if $y=0$ at both ends of a streamline, then the
corresponding axisymmetric surface can serve as the wetted boundary of a motionless
body or as a part of such boundary (see fig.~\ref{fig:cont2}, where the first of
these options is realised for the body floating underneath the other one, whereas
the latter body realises the second option). Assertion (iii) of the last proposition
implies that one of the structure's motionless bodies, say, $B_N$ can be taken so
that $\varphi_m$ is harmonic in $W$; that is, the singularity of $\varphi_m$ is
separated from the water domain by the wetted surface $S_N$ coinciding with a
streamline surrounding the singularity. The next step is to prove the following
assertion.

\vspace{1mm}

\noindent {\sc Proposition 4.2.} {\it For any given positive integer $M$ there
exists $m_*(M)$ such that the function $\psi_m (\nu |\vec{x}|, \nu y)$ has $M-1$
local extrema on $\{\nu |\vec{x}|\in(0, j_{1,M}),\nu y=0\}$ provided $m \geq
m_*(M)$. The points, where these extrema are attained, tend to $(j_{0,l},0)$ as
$m\to\infty$; here $j_{0,l}$, $l=1,\ldots,M$, are zeros of the Bessel function
$J_0$.}

\vspace{1mm}

\noindent {\it Proof.} Since extrema of $\psi_m$ and $c\psi_m$, where $c$ is a
non-zero constant, are attained at the same points, we consider
\begin{gather}
 \frac{\psi_m (\nu |\vec{x}|, \nu y)}{Y_1 (j_{1,m})} = - \pi^2 \nu |\vec{x}| \, \E{\nu y}
 J_1 (\nu |\vec{x}|) + \Lambda(\nu|\vec{x}|,\nu y),
 \label{eq:psi/Y}
\\ \mbox{where} \ \Lambda(\nu|\vec{x}|,\nu y)=\frac{2 \, \nu |\vec{x}|}{Y_1
(j_{1,m})} \int_0^\infty
 \bigl[k \sin (k \nu y) - \cos (k \nu y)\bigr] I_1
 (k \, \nu |\vec{x}|) K_1 (k \, j_{1,m}) \frac{k^2\, \D k}{k^2 + 1} 
\notag
\end{gather}
according to (\ref{eq:psi0<}). Let us show that $\Lambda(\nu|\vec{x}|,0)$ and
$\nabla_{|\vec{x}|,y}\Lambda(\nu|\vec{x}|,\nu y)|_{y=0}$ tend to zero as $m \to
\infty$, uniformly for $\nu |\vec{x}| \in [0, j_{1,M}]$.

We have that
\[ 
|\Lambda(\nu|\vec{x}|,0)|\leq \frac{\nu |\vec{x}|}{|Y_1 (j_{1,m})|}
\int_0^\infty k I_1 (k \, \nu |\vec{x}|) K_1 (k \, j_{1,m}) \, \D k
\]
because $I_1$ and $K_1$ are positive functions. Moreover, the right-hand side is
equal to
\begin{equation}
 \frac{(\nu |\vec{x}|)^2}{j_{1,m} |Y_1 (j_{1,m})| \left[ j_{1,m}^2 - (\nu |\vec{x}|)^2
 \right]} \quad \mbox{for} \ m \geq M+1 \ \mbox{and} \ \nu |\vec{x}| \in [0, j_{1,M}] ,
 \label{eq:m_to_inf}
\end{equation} 
which is a consequence of formula 1.12.4.2 in \cite{PBM}:
\begin{equation}
 \int_0^\infty k I_\mu (a k) K_\mu (b k) \, \D k = \frac{\left( a b^{-1} \right)^\mu}
 {b^2 - a^2} \quad \mbox{for} \ \mu > -1 .
 \label{eq:pbm}
\end{equation}
Note that formulae 9.5.12 and 9.2.2 in \cite{AS} give
\begin{equation}
 j_{1,m} = \pi \left( m + \frac{1}{4} \right) + O ( m^{-1} ) \  \mbox{and} \
Y_1 (j_{1,m}) = (-1)^{m+1} \sqrt{\frac{2}{\pi^2 m}} + O ( m^{-3/2} ) \ \ \mbox{as} \
m \to \infty .
 \label{eq:m_large}
\end{equation}
Since $M$ is fixed, (\ref{eq:m_to_inf}) implies that 
\[ \max\{| \Lambda(\nu|\vec{x}|,
0)| : \nu |\vec{x}| \in [0, j_{1,M}]\}=O\bigl(m^{-5/2}\bigr) \ \mbox{as} \ m \to
\infty .
\]

Using formula 9.6.28 in \cite{AS}, we write
\[ 
\partial_{|\vec{x}|}\Lambda(\nu|\vec{x}|,0)=
\frac{2\nu^2|\vec{x}|}{Y_1 (j_{1,m})} \int_0^\infty I_0 (k \, \nu |\vec{x}|) K_1 
(k \, j_{1,m}) \frac{k^3 \, \D k}{k^2 + 1} \, .
\]
Hence
\begin{eqnarray*}
&& |\partial_{|\vec{x}|}\Lambda(\nu|\vec{x}|,0)|\leq \frac{\nu^2|\vec{x}|} {|Y_1
(j_{1,m})|} \int_0^\infty k^2 I_0 (k \, \nu |\vec{x}|) K_1 (k \, j_{1,m}) \, \D k \\
&& \ \ \ \ \ \ \ \ \ \ \ \ \ \ \ \ \ \ \ \ = \frac{2\nu^2|\vec{x}| j_{1,m}}{|Y_1
(j_{1,m})| \left[ j_{1,m}^2 - (\nu |\vec{x}|)^2 \right]^2} \, .
\end{eqnarray*}
To get the last equality we differentiate (\ref{eq:pbm}), where $\mu = 0$, with
respect to $b$ and use the identity $K_0' (z) = - K_1 (z)$. Now from
(\ref{eq:m_large}) we obtain that
\[ \max\{|\partial_{|\vec{x}|}\Lambda(\nu|\vec{x}|,0)|:\nu |\vec{x}| \in [0,
j_{1,M+1}]\}=\MF{O}\bigl(m^{-5/2}\bigr) \quad \mbox{as} \ m \to \infty .
\]

Furthermore, we have
\[ 
\partial_{y}\Lambda(\nu|\vec{x}|,\nu y)\bigr|_{y=0} = \frac{2\nu^2|\vec{x}|}{Y_1
(j_{1,m})} \int_0^\infty I_1 (k \, \nu |\vec{x}|) K_1 (k \, j_{1,m}) \frac{k^4 \, \D
k}{k^2 + 1} 
\]
and
\begin{eqnarray*}
&& \bigl|\partial_{y}\Lambda(\nu|\vec{x}|,\nu y)\bigr|_{y=0} \leq
\frac{\nu^2|\vec{x}|}{|Y_1 (j_{1,m})|} \int_0^\infty\!\! k^3 I_1 (k \, \nu
|\vec{x}|) K_1 (k \, j_{1,m}) \D k \\ && \ \ \ \ \ \ \ \ \ \ \ \ \ \ \ \ \ \ \ \ \ \
\ \ \ = \frac{8\nu^3|\vec{x}|^2 j_{1,m}}{|Y_1 (j_{1,m})|\! \left[ j_{1,m}^2 - (\nu
|\vec{x}|)^2 \right]^3} \, .
\end{eqnarray*}
Again, the last equality is obtained by differentiation with respect to $a$ and $b$
of (\ref{eq:pbm}) with $\mu = 0$ and using that $K_0' (z) = - K_1 (z)$ and $I_0' (z)
= I_1 (z)$. Then (\ref{eq:m_large}) yields that
\[ \max\{|\partial_{y}\Lambda(\nu|\vec{x}|,\nu y)|:\nu |\vec{x}| \in [0,
j_{1,M}],\,y=0\} = \MF{O} \bigl(m^{-9/2}\bigr) \quad \mbox{as} \ m \to \infty .
\]

It is easy to observe that the first term in the right-hand side of (\ref{eq:psi/Y})
considered in $\overline{Q}$ attains its local extrema on the free surface. Since
$(zJ_1(z))'_z=zJ_0(z)$ \cite[9.1.30]{AS}, the extrema are attained at $(j_{0,l},0)$,
$l=1,2,...$. The sign of these extrema alternates being strictly positive at maxima
(negative at minima) and at these points the $y$-derivative is strictly positive
(negative). It is clear that $(j_{0,l},0) \in \Upsilon=\{\nu|\vec{x}| \in (0,
j_{1,M}),\nu y=0\}$ for all $l=1,2,...,M-1$.

In view of asymptotic estimates obtained for $\Lambda(\nu|\vec{x}|,0)$,
$\nabla_{|\vec{x}|,y}\Lambda(\nu|\vec{x}|,\nu y)|_{y=0}$ as $m \to \infty$ we see
that the contribution of the second term in the right-hand side of (\ref{eq:psi/Y})
is negligible. Therefore, the behaviour of $\psi_m (\nu |\vec{x}|, \nu y)$ on
$\Upsilon$ and in $Q$ near $\Upsilon$ is the same as that of the first term in the
right-hand side of (\ref{eq:psi/Y}). There are $M-1$ points of extrema of $\psi_m
(\nu |\vec{x}|, \nu y)$ on $\Upsilon$ (alternating in sign, strictly positive
maximums and negative minimums), and the points of extrema approach $(j_{0,l},0)$
($l=1,2,...,M-1$) as $m\to\infty$.

A consequence of Propositions 4.1 and 4.2 is the following assertion.

\vspace{1mm}

\noindent {\sc Theorem 4.3.} {\it For every $\omega > 0$ and every integer $N \geq
2$ there exists a motionless structure such that the subsidiary conditions are
fulfilled for each of the structure's $N$ bodies; that is, the structure floats
freely. Moreover, it traps the mode $(\varphi_m, \bm{0}, \dots, \bm{0})$ of
frequency $\omega$; here $\bm{0}$\,---\,the zero element of $\RR^6$\,---\,is
repeated $N$ times.}

\vspace{1mm}

\noindent {\it Proof.} By Proposition 4.2, there exists a sufficiently large $m$
such that $\psi_m (\nu |\vec{x}|, \nu y)$ has $N-1$ local extrema on the interval
$(0, j_{1,N})$ of the free surface. Locally, each of the extrema defines a family of
streamlines enclosing the extrema point with their end-points on the $\nu
|\vec{x}|$-axis. By Proposition 4.1 (iii), there exists one more family of
streamlines with $y=0$ at both end-points of a streamline lying on each side of the
singularity of $\psi_m$ (and $\phi_m$ as well). See fig.~\ref{fig:cont2}, for
examples of streamlines of both types. Choosing a single streamline from each of the
described families and complementing every of chosen streamlines in the same way as
that in the middle of fig.~\ref{fig:cont2} (it is clear that there are infinitely
many other ways to do this), we obtain $N$ axisymmetric bodies.

Using a properly chosen axisymmetric density distribution within each of these $N$
bodies, we get that all subsidiary conditions concerning the bodies' equilibrium are
fulfilled (see again fig.~\ref{fig:cont2}). Indeed, the position of the centre of
mass of each body is on the $y$-axis and can be made arbitrarily close to the level
of its lowest point. Furthermore, every matrix $\widehat{\bm{K}}_0^{(k)}$,
$k=1,\dots,N$, (see (\ref{eq:KK'})) is a diagonal matrix with positive elements in
the limit, and so $\widehat{\bm{K}}_0^{(k)}$ is positive definite when the centre of
mass is sufficiently close to its lowest level.

It remains to show that $(\varphi_m, \bm{0}, \dots, \bm{0})$ is a trapped-mode
solution of problem (\ref{eq:1})--(\ref{eq:5}) in the case of the structure
constructed above. Since $\varphi_m$ satisfies relations (\ref{eq:1}) and
(\ref{eq:2}) and the homogeneous condition (\ref{eq:4}) holds for it on every
constructed $S_k$, we have to verify $6 N$ equations (\ref{eq:5}) which take the
following form
\begin{equation}
\begin{gathered}
 \int_{S_k} \varphi_m\, \partial_{\vec{n}} y \, \D{}s = 0 , 
 \quad \int_{S_k} \varphi_m\, \partial_{\vec{n}} x_i \, \D{}s = 0 ,
 \quad \int_{S_k} \varphi_m\, ( x_2 \partial_{\vec{n}} x_1 -
 x_1 \partial_{\vec{n}} x_2 ) \, \D{}s = 0 , \\ \int_{S_k} \varphi_m \Bigl[
 \bigl( y - y^{(k)}_0 \bigr) \partial_{\vec{n}} x_i - x_i \partial_{\vec{n}} \bigl(
 y - y^{(k)}_0 \bigr) \Bigr] \, \D{}s = 0 , \ \ i = 1,2 ,
\end{gathered}
\label{eq:intS}
\end{equation}
for each of $N$ bodies. We immediately see that the last five of these equalities
are valid because every $S_k$ is axisymmetric, $\varphi_m$ depends on $|\vec{x}|$
and $y$, whereas the second factors in the integrands have the following properties:

\vspace{1mm}

\noindent $\bullet$ $\partial_{\vec{n}} x_i$ and $\bigl( y - y^{(k)}_0 \bigr)
\partial_{\vec{n}} x_i - x_i \partial_{\vec{n}} \bigl( y - y^{(k)}_0 \bigr)$ are odd
functions of the variable $x_i$;\\[1.5mm] $\bullet$ $x_2 \partial_{\vec{n}} x_1 -
x_1 \partial_{\vec{n}} x_2$ is an odd function of both variables $x_1$ and $x_2$.

\vspace{1mm}

Let us show that the first equality (\ref{eq:intS}) holds for every $S_k$, for which
purpose we apply the second Green's identity. First, we check the equality for $S_N$
(we recall that this surface separates the singularity of $\varphi_m$ from the water
domain) and write the following identity
\[ 0 = \int_{\partial ((\RR_-^3 \setminus B_N) \cap C_{b,d})} \bigl( \varphi_m\,
\partial_{\vec{n}} \Yfunc_\nu - \Yfunc_\nu\, \partial_{\vec{n}} \varphi_m \bigr)\, \D
s .
\]
Here $\Yfunc_\nu = y + \nu^{-1}$ and $C_{b,d} = \{ (\vec{x}, y) : \, |\vec{x}| < b ,
\, -d < y < 0 \}$ is a truncated cylinder and $b,\,d > 0$ are taken so that $S_N
\subset C_{b,d}$. Since both functions in this identity are harmonic in $\RR_-^3
\setminus B_N$, the boundary condition (\ref{eq:sing}) yields that
\[ - \int_{S_N} \varphi_m \, \partial_{\vec{n}} y \, \D{}s = \int_{(\partial C_{b,d})
 \setminus \partial\RR_-^3} \bigl( \varphi_m\, \partial_{\vec{n}} y - \Yfunc_\nu \,
 \partial_{\vec{n}} \varphi_m \bigr)\, \D s .
\]
Let us fix $b$ and pass to the limit as $d \to +\infty$ in the last integral, which
we split into the sum of two integrals: one over the bottom $\partial C_{b,d} \cap
\{y=-d\}$, and the other over the lateral surface $\{|\vec{x}| = b , \, -d < y < 0
\}$. The first integral tends to zero because $\varphi_m$ satisfies the same
estimate as $R$ in formula (\ref{eq:uas}). Therefore, we get
\begin{multline}
 \int_{S_N} \varphi_m \, \partial_{\vec{n}} y \, \D{}s = 2 \pi b \lim_{d \to +\infty} 
 \int_{-d}^0 \Yfunc_\nu (y) \, \partial_{|\vec{x}|} \varphi_m (\nu |\vec{x}|, \nu y)
 \Bigl|_{|\vec{x}|=b} \, \D y \\{} = - 2 \pi \nu^{-1} \lim_{d \to +\infty} 
 \int_{-d}^0 \Yfunc_\nu (y)  \partial_y \psi_m (\nu b, \nu y) \, \D y .
\label{eq:limit}
\end{multline}
Here the axisymmetric behaviour of $\varphi_m$ is taken into account in the second
expression, whereas the last equality is a consequence of the first equation
(\ref{eq:ps}).

In order to show that the limit is equal to zero in (\ref{eq:limit}), we substitute
the expressions for $\Yfunc_\nu$ and $\partial_y \psi_m$ into the last integral and
obtain
\begin{equation}
 -2\, \nu b \int_0^\infty I_1 (k\nu r_m) K_1 (k \nu b)
 \frac{k^3\, \D k}{k^2 + 1} \int_{-d}^0 \left( y + \nu^{-1} \right) (k \cos k \nu y
 + \sin k \nu y) \, \D y 
\label{eq:Ae}
\end{equation}
after changing the order of integration. (Indeed, the inequality $b > r_m$ and the
asymptotic formulae for $I_1 (z)$ and $K_1 (z)$ as $z \to 0$ and $z \to +\infty$
(see, for example, \cite{AS}, 9.6.7--9.6.9, 9.7.1, 9.7.2) imply that the double
integral the absolutely convergent.) Since the inner integral is equal to
$\nu^{-2}\left( 1+ k^{-2} \right) \sin k\nu d - d \nu^{-1} \left( \sin k\nu d +
k^{-1} \cos k\nu d \right)$, we get that the contribution of the first term into
(\ref{eq:Ae}) is equal to
\[ -2\, \nu^{-1} b \int_0^\infty k \, I_1 (k \nu r_m) K_1 (k \nu b) 
 \, \sin k \nu d \, \D k 
\]
tending to zero as $d \to +\infty$ by the Riemann--Lebesgue lemma. It remains to
consider the contribution of the second term, which takes the form
\begin{equation}
 \!\!\!\nu^{-1} \int_0^\infty \left( \sin k \nu d - k \cos k \nu d +
 \frac{\sin k \nu d}{\nu d} \right) f (k) \, \D{}k 
 \label{eq:Asp}
\end{equation}
after integration by parts; here
\begin{multline*}
 f (k) = \Bigl\{ I_1 (k \nu r_m) K_1 (k \nu b) \frac{2}{k(k^2 + 1)}
 + \frac{\nu r_m}{2} K_1 (k\nu r) \bigl[ I_0 (k \nu r_m) + 
 I_2 (k \nu r_m) \bigr]\\{} - \frac{\nu b}{2} I_1 (k \nu r_m)
 \bigl[ K_0 (k \nu b) + K_2 (k \nu b) \bigr] \Bigr\} \frac{k^2}{k^2 + 1}.
\end{multline*}
Simple analysis shows that $f$ is integrable, and so (\ref{eq:Asp}) also tends to
zero as $d \to +\infty$ by the Riemann--Lebesgue lemma.

To show that the first equality (\ref{eq:intS}) holds for every $S_k$ with $k \neq
N$ we use the following identity
\[ 0 = \int_{\partial ((\RR_-^3 \setminus \overline{B_N \cup B_k}) \cap C_{b,d})} 
\bigl( \varphi_m \, \partial_{\vec{n}} \Yfunc_\nu - \Yfunc_\nu\, \partial_{\vec{n}}
\varphi_m \bigr)\, \D s .
\]
In the same way as above, this reduces to
\[ - \int_{S_N \cup S_k} \varphi_m \, \partial_{\vec{n}} y \, \D{}s = - \left( 
\int_{S_N} + \int_{S_k} \right) = \int_{(\partial C_{b,d}) \setminus
\partial\RR_-^3} \bigl( \varphi_m\, \partial_{\vec{n}} y - \Yfunc_\nu \,
\partial_{\vec{n}} \varphi_m \bigr)\, \D s .
\]
Since the integral over $S_N$ is equal to zero, the above considerations yield that
the same is true for the integral over $S_k$ with arbitrary $k \neq N$. The proof is
complete.

\subsection{Modified stream functions and \\ heaving trapping structures}

Let us turn to constructing a freely floating trapping structure all bodies of
which are in heave motion with the same amplitude of vertical oscillations; that
is,
\begin{equation}
\vec{\chi}^{(1)}_* = \dots = \vec{\chi}^{(N)}_* = \vec{\chi}_H =
(0,0,0,H,0,0)^\transp , 
\label{eq:chi_H}
\end{equation}
where $H$ is a (sufficiently small) positive constant. For this purpose we
modify the method presented in \S\S~4.2 and 4.3. Namely, we require a structure
to be formed by bodies whose wetted surfaces $\{S_k\}_{k=1}^N$ are level lines
of the form
\begin{equation}
\psi_m^{(H)} (\nu |\vec{x}|, \nu y) = v = \mathrm{const} , \ \ \mbox{where} \ \
\psi_m^{(H)} (\nu |\vec{x}|, \nu y) = \psi_m (\nu |\vec{x}|, \nu y) - \frac{H}{2}
(\nu |\vec{x}|)^2 .
\label{eq:psi_H}
\end{equation}
Indeed, relations (\ref{eq:ps}) imply that
\[ \partial_{\vec{n}}\kern-0.25pt (\varphi_m - H \nu y) = 0 \quad \mbox{on every such}
\ S_k ,
\]
which is equivalent to the Neumann condition (\ref{eq:4}) describing the heave
motion. Similarly to Propositions 4.1 and 4.2, one obtains the following two
propositions illustrated in fig.~4. The existence of the constants $v = v_k$
delivering trapping structures will be shown below.

\vspace{1mm}

\noindent {\sc Proposition 4.4.}~{\it Let $H$ be sufficiently small, then the
following three assertions hold.} \\[1mm]
 (i) {\it Level lines $(\ref{eq:psi_H})$ are smooth curves in\/ $Q$; their 
end-points belong to\/ $\partial Q \cup \{\infty\}$.} \\[1mm]
 (ii) {\it A level line emanates from every point on the half-axis\/ $\{ \nu |\vec{x}| 
> 0 , \, \nu y = 0 \}$, except for the points, where\/ $\psi_m^{(H)} (\nu
|\vec{x}|, \nu y)$ attains its local extrema, and the point\/ $(\nu r_m, 0)$.}
\\[1mm]
 (iii) {\it For every $m \geq 1$, all positive\/ $\nu$ and all sufficiently large 
values of $v_N$ $($in $(\ref{eq:psi_H}),$ the same notation is used as at the
beginning of\/ \S\,$4.3)$ there exists a level line such that $y=0$ at both its ends
and the point\/ $(\nu r_m , 0)$ belongs to the segment connecting these end-points.}

\vspace{1mm}

\noindent {\sc Proposition 4.5.} {\it For any positive integer $M$ there exist
$m_*(M)$ and $H_*(M)$ such that the stream function $\psi_m^{(H)} (\nu |\vec{x}|,
\nu y)$ has $M-1$ local extrema on 
\[ \{\nu |\vec{x}|\in(0, j_{1,M}), \nu y=0\}
\]
provided $m\geq m_*$ and $H\leq H_*$.}

\begin{figure}[t]
\begin{center}
 \SetLabels
 \L (-0.08*0.94) (a)\\
 \L (-0.08*0.61) (b)\\
 \L (0.93*0.705) {\small $\nu|\vec{x}|$}\\
 \L (0.93*0.01) {\small $\nu|\vec{x}|$}\\
 \L (-0.02*0.45) {\small $\nu y$}\\
 \L (0.47*0.9) {\small $\psi_2^{(H)} (\nu |\vec{x}|,0)$}\\
 \endSetLabels
\leavevmode\kern2mm\AffixLabels{\includegraphics[width=96mm]{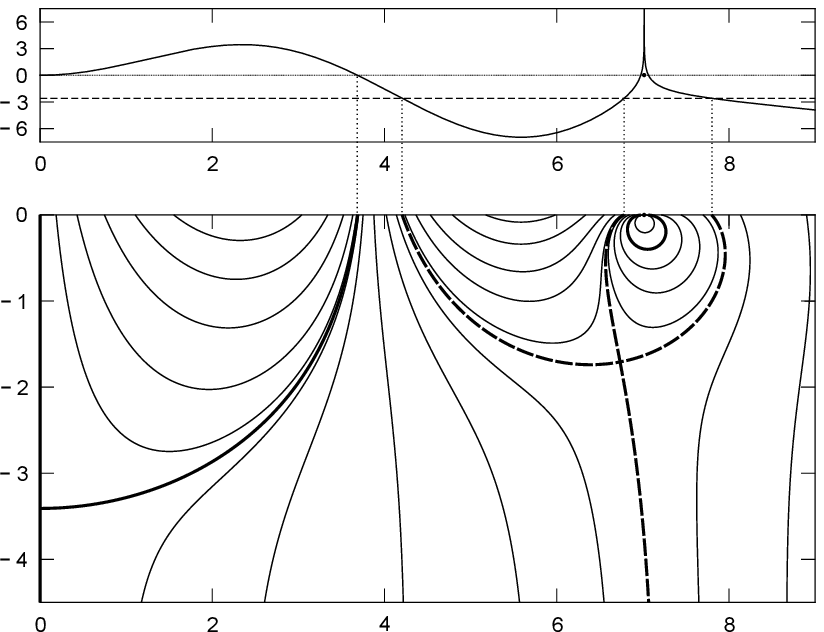}}
 \end{center}
 \vspace{-4.5mm}
 \caption{Let $H=0.1$, then: the trace $\psi_2^{(H)} (\nu |\vec{x}|, 0)$ is plotted
 in (a); level lines of $\psi_2^{(H)} (\nu |\vec{x}|, \nu y) = v$ are plotted
 in (b) for various values of $v$. Solid bold lines correspond to $v=0$ (nodal lines); 
 the dashed lines in (a) and (b) correspond to $v \approx -2.590$ (the level of the
 right stagnation point).}
 \label{fig:cont2H}
\vspace{-4mm}
\end{figure} 

\vspace{1mm}

\noindent {\sc Remark 4.6.} The level line corresponding to $v$ and going to
infinity asymptotes the vertical line $\nu |\vec{x}| = \sqrt{2 v / H}$ (in fig.~4,
such lines are located between the $\nu y$-axis and the dashed line going to
infinity and to the right of the latter line).

\vspace{1mm}

A consequence of Propositions 4.4 and 4.5 is the following theorem.

\vspace{1mm}

\noindent {\sc Theorem 4.7.} {\it For every $\omega > 0$ and every integer $N \geq
2$ there exists a heaving structure such that the wetted surfaces of its bodies are
given by level lines of $\psi_m^{(H)}$ with sufficiently small $H$ and properly
chosen values $v_k$, $k=1, \dots,N$ $($see fig.~$4)$. For each of $N$ bodies the
subsidiary conditions guarantee its equilibrium; that is, the structure floats
freely. Moreover, it traps the mode $(\varphi_m, \vec{\chi}_H, \dots,
\vec{\chi}_H)$.}

\vspace{1mm}

\noindent {\it Proof.} Applying Propositions 4.4 and 4.5 in the same way as
Propositions 4.1 and~4.2 were applied in the proof of Theorem 4.3, we obtain $N$
axisymmetric bodies by choosing the level lines $\psi_m^{(H)} (\nu |\vec{x}|, \nu y)
= v_k$ so that the values $\{ v_k \}_{k=2}^N$ are close (but not equal) to the
extrema values of $\psi_m^{(H)} (\nu |\vec{x}|, 0)$ on the interval $(0, j_{1,N})$;
the level line $\psi_m^{(H)} (\nu |\vec{x}|, \nu y) = v_N$, where $v_N$ is
sufficiently large, gives the wetted surface $S_N$ of $\widehat{B}_N$ (the
cross-section of this body must be the rightmost in fig.~4).

All subsidiary conditions hold for these bodies provided axisymmetric density
distributions are properly chosen within each body. As in Theorem~4.3, the velocity
potential $\varphi_m$ satisfies relations (\ref{eq:1}) and (\ref{eq:2}), whereas the
boundary condition (\ref{eq:4}) holds on every $S_k$ in view of the way how these
surfaces are constructed using the displacement vectors (\ref{eq:chi_H}). It remains
to verify $6 N$ equations (\ref{eq:5}) of which $5 N$ take the same form as
equalities (\ref{eq:intS}) with the exception of the first one. The latter is as
follows:
\begin{equation}
\nu H I^{\widehat{B}_k} = - \int_{S_k} \varphi_m \, \partial_{\vec{n}} y \,
\D{}s + H I^{D_k} , \quad k=1,\dots,N . \label{eq:last}
\end{equation}
As in the proof of Theorem 4.3, we begin with the following second Green's
identity
\[ 0 = \int_{\partial ((\RR_-^3 \setminus \overline{B_N}) \cap C_{b,d})} 
\bigl( \varphi_m \, \partial_{\vec{n}} \Yfunc_\nu - \Yfunc_\nu\, \partial_{\vec{n}}
\varphi_m \bigr)\, \D s ,
\]
to which we apply the same considerations based on the boundary conditions, the
behaviour of $\varphi_m$ as $y \to - \infty$ and the Riemann--Lebesgue lemma
with $d \to + \infty$. However, now we obtain that
\[ \int_{S_N} \varphi_m \, \partial_{\vec{n}} y \, \D{}s = H \nu \int_{S_N} 
(y + \nu^{-1}) \, \partial_{\vec{n}} y \, \D{}s  = - H \nu \int_{B_N} \D{}\vec{x}
\D{}y + H I^{D_N} .
\]
Substituting this into (\ref{eq:last}) with $k=N$, the latter equality reduces to
Archimedes' law for $\widehat{B}_N$, and so is true.

Then the same procedure yields the result for $k \neq N$, but we have to apply the
second Green's identity in $(\RR^3_- \setminus \overline{B_N \cup B_k}) \cap
C_{b,d}$ and to take into account the fact obtained on the previous step as well as
Archimedes' law for $\widehat{B}_k$. The proof is complete.

\vspace{1mm}

There are various ways to construct trapping structures using level lines
(\ref{eq:psi_H}), in particular, level lines plotted in fig.~4 allow us to obtain
four different types of structures with $N = 2,3$, one of which is similar to that
shown in fig.~3.

\begin{figure}[t]
\begin{center}
 \SetLabels
 \L (-0.1*0.98) (a)\\
 \L (-0.1*0.83) (b)\\
 \L (0.9*0.87) {\small $\nu|\vec{x}|$}\\
 \L (0.9*0.475) {\small $\nu|\vec{x}|$}\\
 \L (-0.04*0.77) {\small $\nu y$}\\
 \L (0.12*0.965) {\small $\psi_1^{(H)} (\nu |\vec{x}|, 0)$}\\
 \L (0.04*0.843) {\small Rigid}\\
 \L (0.04*0.818) {\small shell}\\
 \L (0.487*0.859) {\small Air}\\
 \L (0.69*0.818) {\small Ballast}\\
 \L (-0.1*0.435) (c)\\
 \L (-0.1*0.275) (d)\\
 \L (0.9*0.295) {\small $\nu|\vec{x}|$}\\
 \L (0.9*-0.01) {\small $\nu|\vec{x}|$}\\
 \L (-0.04*0.235) {\small $\nu y$}\\
 \L (0.12*0.4) {\small $\psi_1 (\nu |\vec{x}|, 0)$}\\
 \endSetLabels

\leavevmode\kern6mm\AffixLabels{\includegraphics[width=63mm]{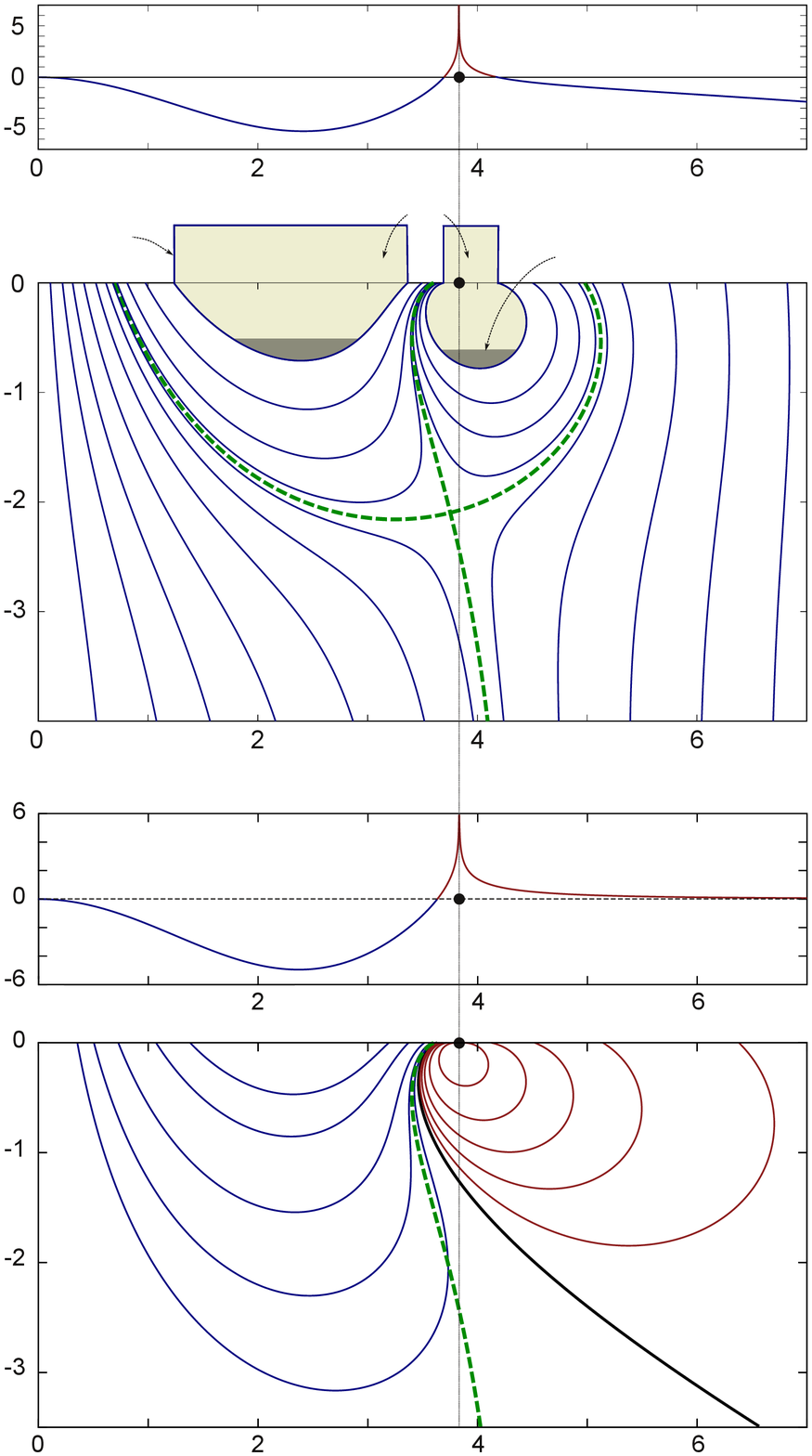}}
 \end{center}
 \vspace{-2mm}
 \caption{\leftskip=2mm (a) The trace $\psi_1^{(H)} (\nu |\vec{x}|,0)$ with 
 $H = 0.1$. (b) Streamlines $\psi_1^{(H)} (\nu |\vec{x}|, \nu y) = v$ are
 plotted for $H = 0.1$ and various $v \leq 0$; the nodal line ($v=0$) serves 
 as $S_2$ (the wetted boundary of the right body); straight segments connect 
 wetted surfaces above the free surface to form two bodies; two dashed lines 
 ($v \approx -0.9464$) separate four different families of level lines. (c) The 
 trace $\psi_1 (\nu |\vec{x}|, 0)$. (d) The dashed line is taken from (b); 
 streamlines $\psi_1 (\nu |\vec{x}|, \nu y) = v$ are plotted for various 
 values of $v$; the nodal line ($v=0$) is bold.}
 \label{fig:cont4}
\vspace{-4.5mm}
\end{figure}

\subsection{Trapping structures consisting of two bodies}

In the simplest case $N=2$, we describe the whole set of trapping structures
expressible in terms of $\psi_1$ and $\psi_1^{(H)}$ with sufficiently small $H$. We
begin with the following proposition proved in \cite{KM1}, pp.~153 and 154, and
illustrated in fig.~5(c).

\vspace{1mm}

\noindent {\sc Proposition 4.8.} {\it For any $\nu > 0$ the trace\/ $\psi_1 (\nu
|\vec{x}|, 0)$ has the following properties. It vanishes at $\nu |\vec{x}| = 0$,
tends to $+\infty$ as\/ $\nu |\vec{x}| \to j_{1,1} \pm 0$, tends to $0$ as\/ $\nu
|\vec{x}| \to +\infty$, has exactly one zero on\/ $(0, j_{1,1})$, say, $\nu
|\mathring{\vec{x}}|$ and exactly one extremum on\/ $(0, j_{1,1})$, namely, the
negative minimum $\breve{M}$ attained at a certain\/ $\nu |\breve{\vec{x}}| < \nu
|\mathring{\vec{x}}|$.}

\vspace{1mm}

According to this proposition streamlines exist only for $v > \breve{M}$ and only
one streamline corresponds to each $v \in (\breve{M}, 0)$ and to each $v > 0$ (see
fig.~5(d)). The nodal streamline emanating from $(\nu |\mathring{\vec{x}}| , 0)$
separates streamlines that correspond to positive and negative levels of $\psi_1
(\nu |\vec{x}|, \nu y)$. It is straightforward to show that the nodal line does not
intersect the $\nu y$-axis and goes to infinity.

Proposition 4.8 and the definition of $\psi_1^{(H)}$ yield the following corollary
illustrated in fig.~5(a).

\vspace{1mm}

\noindent {\sc Corollary 4.9.} {\it For any $\nu > 0$ and sufficiently small $H$
the trace\/ $\psi_1^{(H)} (\nu |\vec{x}|, 0)$ has the following properties. It
vanishes at $\nu |\vec{x}| = 0$, tends to $+\infty$ as\/ $\nu |\vec{x}| \to j_{1,1}
\pm 0$, tends to $-\infty$ as\/ $\nu |\vec{x}| \to +\infty$, has exactly one zero
on\/ $(0, j_{1,1})$, say, $\nu |\mathring{\vec{x}}^{(H)}|$ and exactly one extremum
on\/ $(0, j_{1,1})$, namely, the negative minimum $\breve{M}^{(H)}$ attained at a
certain\/ $\nu |\breve{\vec{x}}^{(H)}| < \nu |\mathring{\vec{x}}^{(H)}|$.}

\vspace{1mm}

According to this corollary level lines such that $y=0$ at both their end-points
exist only for $v > \breve{M}^{(H)}$ and only one such level line corresponds to
each $v > \breve{M}^{(H)}$ (see fig.~5(d)). The presence of the second negative term
in the definition of $\psi_1^{(H)}$ has the following consequence. Instead of the
nodal streamline separating two families of streamlines defined by $\psi_1$, the
same role for level lines of $\psi_1^{(H)}$ with sufficiently small $H$ is played by
the branch that corresponds to a certain critical negative level $v$ (for $H = 0.1$
this level is $\approx -0.9464$) and goes to infinity. The second branch of the
critical level has $y=0$ at its both end-points, thus separating two families of
level lines having $y=0$ at both end-points from those going to infinity (see
Remark~4.6). In fig.~5(b), these two branches are shown by dashed lines.

\vspace{1mm}

\noindent {\sc Theorem 4.10.} {\it There exist four types of trapping structures
defined by $\psi_1$ and $\psi_1^{(H)}$ with sufficiently small $H$. Every structure
consists of two bodies and the corresponding trapped mode is either of the following
four: $(\varphi_1, \bm{0}, \bm{0})$, $(\varphi_1, \vec{\chi}_H, \vec{\chi}_H)$,
$(\varphi_1, \bm{0}, \vec{\chi}_H)$, $(\varphi_1, \vec{\chi}_H, \bm{0})$, where
$\vec{\chi}_H$ is defined by $(\ref{eq:chi_H})$.}

\vspace{1mm}

\noindent {\it Proof.} According to Proposition 4.8, there exist two families of
streamlines defined by $\psi_1$; every streamline belonging to the first family
surrounds the singularity of $\varphi_1$, whereas streamlines of the second family
are separated from the former ones by the nodal line of $\psi_1$ (see fig.~5(d)).
Let us take $S_1$ and $S_2$ arbitrarily from different families and complement these
two streamlines by, for example, rectangles in the same way as in fig.~5(b). Then we
obtain a structure of two bodies satisfying all subsidiary conditions provided
appropriate axisymmetric density distributions are chosen. It follows from equations
(\ref{eq:ps}) that the mode $(\varphi_1, \bm{0}, \bm{0})$ is trapped by this
motionless structure floating freely which is guaranteed by our construction.

Considering three other cases we omit for the sake of brevity the words about
complementing $S_1$ and $S_2$ by parts located above the free surface and about a
proper choice of axisymmetric density distributions to satisfy all subsidiary
conditions.

\looseness=-1 The same considerations as above, but using Corollary 4.9 instead of
Proposition 4.8, allow us to take arbitrary level lines $S_1$ and $S_2$ from two
different families defined by $\psi_1^{(H)}$ with sufficiently small $H$ to form a
structure of two bodies (see fig.~5(b)). It follows from equations (\ref{eq:ps})
that the mode $(\varphi_1, \vec{\chi}_H, \vec{\chi}_H)$ is trapped by this heaving
structure floating freely.

To obtain trapping structures consisting of two bodies, one of which is motionless
and the other one is heaving, we use both: streamlines of $\psi_1$ and level lines
of $\psi_1^{(H)}$. Let $S_2$ be an arbitrary streamline of $\psi_1$ surrounding the
singularity. It immediately follows from the definition of $\psi_1^{(H)}$ that the
nodal line of $\psi_1$ lies strictly above the critical branch of $\psi_1^{(H)}$
going to infinity. Therefore, an arbitrary level line of this function can be taken
as $S_2$ provided it lies to the left of the mentioned critical branch and has $y=0$
at its both end-points. Again, equations (\ref{eq:ps}) yield that the mode
$(\varphi_1, \bm{0}, \vec{\chi}_H)$ is trapped by this combined (motionless/heaving)
structure floating freely.

Similarly, a heaving/motionless structure trapping the mode $(\varphi_1,
\vec{\chi}_H, \bm{0})$ consists of an arbitrarily taken level line of $\psi_1^{(H)}$
surrounding the singularity (it is $S_2$ in this case), but as $S_1$ we can take
only any of those streamlines of $\psi_1$ that lie totally to the left of the
critical branch of $\psi_1^{(H)}$ going to infinity. The proof is complete.

\vspace{1mm}

\noindent {\sc Remark 4.11.} It is possible to obtain more complicated
heaving/motionless structures consisting of two bodies (see, for example, figs.~3
and 4) and also trapping structures such that each of their two bodies heaves with
its own sufficiently small amplitude $H_1 \neq H_2$ (see the next section).

\subsection{The general case}

Here we construct general heaving/motionless structures such that they consist of
$N$ freely floating bodies and trap modes of the form
\begin{equation}
\big( \varphi_m, \vec{\chi}^{(1)}_*, \dots, \vec{\chi}^{(N)}_* \big) , \quad
\mbox{where} \ \vec{\chi}^{(k)}_* = (0,0,0, H_k , 0,0)^\transp , \ k = 1,\dots,N ,
\label{eq:combined}
\end{equation}
and $H_k = 0$ ($H_k > 0$) corresponds to the motionless (heaving, respectively)
$k$th body.

\vspace{1mm}

\noindent {\sc Theorem 4.12.}\,{\it For every integer $N \geq 2$, every $\omega >
0$ and every $N$-tuple $(H_1,\dots,H_N)$ of non-negative numbers of which all
positive are sufficiently small there exist a freely floating structure such that it
consists of $N$ bodies and traps the mode $(\ref{eq:combined})$ defined by $\omega$
and $(H_1,\dots,H_N)$.}

\vspace{1mm}

\noindent {\it Proof.} The assertion is already proved in two particular cases:
when all $H_k$ are zeroes or are equal to the same sufficiently small positive
number (see Theorems~4.3 and 4.7, respectively).

Since construction of a structure so that condition (\ref{eq:4}) holds on every its
wetted surface $S_k$, $k = 1,\dots,N$, is the main point of the proof, we
concentrate only on it. Indeed, there is no need to verify relations (\ref{eq:1})
and (\ref{eq:2}) because we use $\varphi_m$, whereas to show that (\ref{eq:5}) holds
one has to apply the method used for this purpose in the proofs of Theorems~4.3 and
4.7. Moreover, it is always possible to choose the density distributions $\{ \rho_k
\}_{k=1}^N$ so that all subsidiary conditions are fulfilled for each body.

According to Propositions 4.2 and 4.5 the functions 
\[ \psi_m (\nu |\vec{x}|, \nu y) \quad \mbox{and} \quad \psi_m^{(H_k)} (\nu |\vec{x}|, 
\nu y)
\]
have $N-1$ local extrema on the interval $(0, j_{1,N})$ of the free surface
provided $m$ is sufficiently large and $H_k$ is sufficiently small. Therefore, one
obtains a motionless (heaving) surface $S_k$ for every $k = 1,\dots,N-1$ by taking a
streamline (level line, respectively) corresponding to a certain value close to the
extremum value of $\psi_m$ $\big( \psi_m^{(H_k)}, \mbox{respectively}\big)$ on the
interval whose number is $k$ counting from the origin. Chosen in this way, surfaces
$S_k$ do not overlap provided they are sufficiently small and separated by large
enough spacings. Indeed, this is a consequence of the fact that the extrema of
$\psi_m (\nu |\vec{x}|, 0)$ and $\psi_m^{(H_k)} (\nu |\vec{x}|, 0)$ are close to
zeros of $J_0 (\nu |\vec{x}|)$. Finally, they must be complemented by motionless
(heaving) surface $S_N$ defined by a streamline of $\psi_m$ $\big( \mbox{level line
of} \ \psi_m^{(H_k)}, \mbox{respectively}\big)$ corresponding to a sufficiently
large positive value in order to get a small surface not overlapping with $S_{N-1}$.
Complementing $S_1,\dots,S_N$ by, for example, rectangles above the free surface to
obtain closed shells, we complete our construction and the proof as well.

\section{Conclusion}

It has been shown that there exist three-dimensional structures and the
corresponding time-harmonic wave modes with the following properties. Every
structure consists of two or more bodies all of which have the same vertical axis of
symmetry and float freely in infinitely deep water; some of these bodies (may be
none) are motionless, whereas the others (may be none) heave at the frequency of the
wave mode. Thus, the latter is trapped, that is, the coupled motion of the structure
and water at this frequency does not radiate waves to infinity, and so, in the
absence of viscosity, will persist for all time. Such structures and wave modes
exist for all frequencies and the structure's geometry depends on the frequency.

\end{document}